Comments on 'Thermometer Effect: Origin of the Mixed Alkali Effect in Glass Relaxation' by Y.Yu et al. [Phys.Rev.Lett. 122, 095501 (2319)]

Vladimir Belostotsky

Institute of General and Inorganic Chemistry, Russian Academy of Sciensces, Moscow, Russia *

In a recent Letter to this journal Yu et al. [1] presented the results of molecular dynamics (MD) simulations of relaxation processes in mixed alkali (MA) glasses in response to cyclic volumetric stress perturbations. Using Phillips's diffusion-trapping model they attempted to link observed stretched exponential relaxation to specific processes in glasses on atomic level and, on this basis, explain the mixed alkali effect (MAE).

Our main concern is the results of the MD simulations, specifically the observation that *"the relaxation of enthalpy and volume do not show the same trend"*.

Basically, enthalpy is extensive thermodynamics state variable defined as the sum of internal energy and product of pressure and system's volume:

$$H = U + pV \qquad (1)$$

This equation may be rewritten in differential form:

$$dH = dU + d(pV) = dU + pdV + Vdp \qquad (2)$$

Within the isothermal-isobaric ensemble (NPT) used in MD simulations in question extensive thermodynamic variables as to pressure, $p$, and temperature, $T$, are constrained. In the absence of heat transfer ($T$=const) $dU=0$, and at constant pressure $dp=0$, hence $Vdp=0$, therefore

$$dH = pdV \qquad (3)$$

As can be seen, in MD simulations in question volume and enthalpy should relax in unison irrespective of the magnitude of pressure.

Experimental studies of alkali silicate glasses revealed that below $T_g$ cyclic volumetric stress perturbations induce relaxation processes connected with diffusional hops of mobile alkali cations and reorientational motions of non-bridging oxygen (NBO) [2-9]. It is an established fact that in MA glasses magnitude of alkali-motion-related relaxation reaches minimum at equimolar concentrations of alkalis while NBO-related process shows an opposite trend. In the work of Yu et al. such or similar trend in obtained results is not reported.

All the above poses the question of the reliability of MD simulations which limits the significance of the obtained results.

Also, the discussion part of the Letter causes some concerns. As was mentioned above, the only species capable to diffuse in glass at ambient temperature are alkali cations. Authors suggest, however, that the relaxation of compression stress in glass arising from the replacement of smaller cations for larger ones, and tension stress arising from the replacement of larger cations for smaller ones can occur via certain diffusion of local deformations (referred to as excitations) through glass network and their mutual annihilation. This assumption ignores the fact that local stress involves up to several coordination shells around cation-centered polyhedrons [10]. The diffusion of excitations of the size of several coordination shells would demand spatial rearrangement of relatively large volume of network which assumes viscous relaxation behavior. However, at ambient temperature, glass network exhibits brittle-elastic and not viscous response to stress. Moreover, calculated differential stress (approx. 4.7 GPa, see Fig 3d in the Letter in question) cannot, in principle, be absorbed by rigid glass matrix without its damage which in alkali silicate glasses typically occurs via formation of pairs of structural defects, oxygen vacancies and non-bridging oxygen anions [11]. Most vividly this effect manifests itself in glasses subjected to ion exchange in molten salt both for smaller-for-larger and larger-for-smaller cation inter-replacement [12-15]. It is an established fact that introduction of defects in glasses causes their compaction [16-18] because finer fragments can be packed tighter that large. MD simulations confirm that defect-induced compaction of glasses occurs regardless of means of defect introduction [19]. Defect formation and glass compaction as a result of smaller-for-larger and larger-for-smaller cation inter-replacement is, in fact, the real cause of the 'thermometer effect' [20].

Authors speculate, as well, that the *"coexistence of atomic units that are under compression or tension can also explain the decrease in the mobility of the alkali atoms in mixed glasses, which results in minima in conductivity and diffusion coefficients"*. Actually, it is proven that tension stress enhances ionic mobility. The theory indeed predicts cations' mobility reduction under compression stress by restricting the amount of volume available [22,23], however it was established that in mixed alkali glasses ionic mobility decreases by significantly larger factor than the theory predicts [22].

Finally, the authors maintain that *"the structural origin of the MAE [is] still regarded as one of the most challenging unsolved problem in condensed matter science"*, and *"the atomic origin of the MAE itself remains largely unknown"*. However, the MAE problem has already been resolved with the introduction of the defect model for the mixed mobile ion effect [20,21] which provides comprehensive, consistent and generally applicable fundamental explanation for MAE in all its facets and agrees with all experimental facts.

All the above criticism, in a shorter version fit for the comment's framework of the journal, had appeared in the Physical Review Letters [24] along with the authors' reply where Bauchy *et al*.

[25] attempted to completely refute the Comments' concerns and criticism. Their argumentation, however, appears to be not convincing and poses more questions than it offers answers:

_Thermodynamics_ - Bauchy *et al.* [25] argue that in MD simulations in [1] they succeeded to decouple enthalpy relaxation from volume relaxation by "*energy minimization ($T=0$ K)*" and "*imposing a zero average stress ($P=0$)*" which would supposedly exclude the second component ($PV$) in the enthalpy definition ($H=U+PV$) and make enthalpy a function of internal energy, $U$, only.

Firstly, this argument is flawed. It confuses externally imposed **average stress** with internal pressure of the system that is caused by forces of interaction between atoms and is computed as a partial derivative of internal energy with respect to volume at constant temperature

$$P = \left(\frac{\partial U}{\partial V}\right)_T \tag{4}$$

The Letter [1] maintains that simulated MA glasses were "*made of 2991 atoms*", and interactions between atoms were taken into account by "*using the well-established Teter potential*" and "*Coulomb interactions were evaluated by the Ewald summation method*". Unlike an ideal gas, in a system of interacting atoms internal pressure can not be zero by the definition. Therefore in such a system enthalpy relaxation cannot, in principle, be decoupled from volume relaxation.

Secondly, in the Reply [25] authors mention some "*effect of heat that is exchanged with the thermostat*". Certain microscopic "*residual thermal excitations*" of unspecified nature and origin that affect relaxation mechanisms are also mentioned in the Letter [1]. Does it mean that in MD simulations MA glasses act as an engine that transforms energy of mechanical perturbations into heat? So far MA glasses were not known being any kind of heat engine.

Thus, the line of reasoning in the Comment [24], namely that enthalpy relaxation cannot be decoupled from volume relaxation and they must relax in unison, is rigorous and the concerns of the reliability of MD simulations in [1] are quite relevant.

*Atomic mechanism* – The Reply [25] unintentionally attracts reader's attention to the fact that the title of Letter [1] does not seem reflective of its content because MD simulations conducted at T=0 K appear to be unrelated to and therefore irrelevant for elucidation of the thermometer effect and, as was shown in Comment [24], they do not reproduce features of mixed alkali effect, while the discussion section is unrelated to MD simulations.

It is worth recollecting what the thermometer effect is, as it was discovered: When a calibrated liquid-in-glass (mercury in Thuringian MA glass) thermometer was immersed into boiling water ($100^{o}C$ or 373.15 K) and later into melting ice ($0^{o}C$ or 273.15 K), it had shown $-0.5^{o}C$ instead of $0^{o}C$. It is an established fact that zero-point depression is caused by structural rearrangements of MA glass towards its compaction [26]. As for MD simulations in question, they were conducted, as the authors maintain, at a temperature T = 0 K ($-273.15^{o}C$) "*to avoid any thermal contribution to the relaxation subsequently computed*", and mimicked "*relaxation observed in granular materials subjected to vibrations*" [1]. How these MD simulations can in principle be related to the MAE effect in general and to the thermometer effect in particular is hard to comprehend, because MAE in glasses is not observed below 230 K [20,27].

*Internal friction* - Bauchy *et al.* [25] try to scorn some of the references in the Comment [24] by misrepresenting them as unrelated to the topic of their Letter [1]. Specifically, their criticism is directed to the term "internal friction" in silicate glasses that is used in several references. As a matter of fact, the term "*internal friction*" in application to glasses is an obsolete synonym for "mechanical relaxation". Therefore, the Comment [24] properly refers readers to the series of papers that discuss mechanical relaxation of MA glasses subjected to cyclic stress perturbations which is supposedly the topic of the Letter [1].

*Origin of the mixed-alkali effect* – In the Reply Bauchy *et al.* [25] opined that they "*do not think that is fair to claim that the **mixed modifier effect** has been "resolved"*" by the "*defect model for the mixed **mobile** ion effect*" [20,21]. We must completely agree with them because mixed modifier effect (MME) is typically referred to **mixed alkaline earth glasses** where alkaline earth atoms are essentially **immobile**.

Bauchy et al also question the ability of the defect model for the mixed mobile ion effect to explain the MAE in all its facets. Specifically, they point out to the anomalous mechanical properties (positive deviation from linearity of Vickers hardness) exhibited by MA glasses [28, 29]. This facet of the MAE effect, indeed, has not been covered in [20,21], however this does not mean that the defect model is incapable to provide such an elucidation.

Like the thermometer effect, positive deviation from linearity of Vickers hardness of MA glasses can also be attributed to the defect-induced compaction of glasses. Indeed, Vickers hardness, $H_V$, is a parameter characterizing in general way the resistance of a solid to compression. It is related to the average interatomic distance, $R$, by

$$H_V = K/R^n \qquad (5)$$

where the parameter $K$ characterizes the interatomic bond strength, and exponent $n$ depends on the bond type [30]. This relation indicates that Vickers hardness of MA glasses must rise with decreasing average interatomic distance which occurs as a result of the defect-induced compaction.

Again, it must be emphasized that MAE is noticeable on the multiple properties of MA glasses. Therefore, "it is essential for the success of a theory that it agrees, at least qualitatively, with all experimental facts." [31]. The defect model for the mixed mobile ion effect is the only theory of MAE that meets this requirement.